\documentclass[prl,twocolumn,floatfix]{revtex4}
\usepackage{amssymb}
\usepackage{amsmath}
\usepackage{epsfig}
\usepackage{graphicx}

\begin{document}

\title{Density oscillation in highly flattened quantum elliptic rings and
tunable strong dipole radiation}
\author{ S.P. Situ}
\author{ Y.Z. He}
\author{ C.G. Bao$^{\ast }$ }
\affiliation{The State Key Laboratory of Optoelectronic Materials and Technologies, \
Zhongshan University, \ Guangzhou, 510275, P.R. China}

\begin{abstract}
A narrow elliptic ring containing an electron threaded by a magnetic
field $ B $ is studied. When the ring is highly flattened, the
increase of $B$ would lead to a big energy gap between the ground
and excited states, and therefore lead to a strong emission of
dipole photons. \ The photon frequency can be tuned in a wide range
by changing $B$\ and/or the shape of the ellipse. The particle
density is found to oscillate from a pattern of distribution to
another pattern back and forth against $B$. This is a new kind of
Aharonov-Bohm oscillation originating from symmetry breaking and is
different from the usual oscillation of persistent current.
\end{abstract}

\maketitle

$\ast $Corresponding author

\bigskip

It is recognized that micro-devices are important to
micro-techniques. Various kinds of micro-devices, including the
quantum rings,$^{1}$ have been extensively studied theoretically and
experimentally in recent years. Quantum rings are different from
other devices due to their special geometry. A distinguished
phenomenon of the ring is the Aharonov-Bohm (A-B) oscillation of the
ground state energy and persistent current$^{2-5}.$ It is believed
that geometry would affect the properties of small systems.
Therefore, in addition to circular rings, elliptic rings or other
rings subjected to specific topological transformations deserve to
be studied, because new and special properties might be found. \
There have been a number of literatures devoted to elliptic quantum
dots$^{6-9}$ and rings$ ^{10-12}$. \ It was found that the elliptic
rings have two distinguished features. (i) The avoided crossing of
the levels and the suppression of the A-B\ oscillation. (ii) The
appearance of localized states which are related to bound states in
infinite wires with bends.$^{13}$ These\ feature would become more
explicit if the eccentricity is larger and the ring is narrower.

On the other hand, as a micro-device, the optical property is obviously
essential to its application. It is guessed that very narrow rings with a
high eccentricity might have special optical property, this is a point to be
clarified. This paper is dedicated to this topic. It turns out that the
optical properties of a highly flattened narrow ring is greatly different
from a circular ring due to having a tunable energy gap, which would lead to
strong dipole transitions with wave length tunable in a very broad range
(say, from 0.1 to 0.001cm). Besides, a kind of A-B density-oscillation
originating from symmetry breaking was found as reported as follows.

We consider an electron with an effective mass $m^{\ast }$\ confined
on a one-dimensional elliptic ring with a half major axis $r_{ax}$\
and an eccentricity $\varepsilon $. Let us introduce an argument
$\theta $ so that a point $(x,y)$ at the ring is related to $\theta
$ as $x=r_{ax}\cos \theta $ and $y=r_{ay}\sin \theta $, where
$r_{ay}=$\ $r_{ax}\sqrt{1-\varepsilon ^{2}} $ is the half minor
axis. A uniform magnetic field $B$ confined inside a cylinder with
radius $r_{in}$ vertical to the plane of the ring is applied. The
associated vector potential reads
$\mathbf{A}=Br_{in}^{2}\mathbf{t}/2r$, where $\mathbf{t}$ is a
unit\textbf{\ }vector normal to the position vector $ \mathbf{r}$.
Then, the Hamiltonian reads
\begin{eqnarray}
H &=&G/(1-\varepsilon ^{2}\cos ^{2}\theta )[-\frac{d^{2}}{d\theta ^{2}}
-i2\alpha \frac{\sqrt{1-\varepsilon ^{2}}}{(1-\varepsilon ^{2}\sin
^{2}\theta )}\frac{d}{d\theta }  \notag \\
&&+\alpha ^{2}\frac{1-\varepsilon ^{2}\cos ^{2}\theta }{1-\varepsilon
^{2}\sin ^{2}\theta }]
\end{eqnarray}
where \bigskip $G=\hbar ^{2}/(2m^{\ast }r_{ax}^{2}),\ \alpha =\phi /\phi
_{o} $, $\phi =\pi r_{in}^{2}B$ is the flux, $\phi _{o}=hc/e$ is the flux
quantum.

The eigen-states are expanded as $\Psi _{j}=\sum_{k=k_{\min }}^{k_{\max
}}C_{k}^{(j)}e^{ik\theta }$, where $k$ is an integer ranging from $k_{\min }$
to $k_{\max }$, and $j=1,2,\cdot \cdot \cdot $ denotes the ground state, the
second state, and so on. The coefficients $C_{k}^{(j)}$ are obtained via the
diagonalization of $H$. In practice, $B$ takes positive values, $k_{\min
}=-100$ and $k_{\max }=10$. This range of $k$ assures the numerical results
having at least four effective figures. The energy of the $j-th$ state is
\begin{equation}
E_{j}=\langle H\rangle _{j}\equiv \int d\theta (1-\varepsilon ^{2}\cos
^{2}\theta )\Psi _{j}^{\ast }H\Psi _{j}
\end{equation}
where the eigen-state is normalized as
\begin{equation}
1=\int d\theta (1-\varepsilon ^{2}\cos ^{2}\theta )\Psi _{j}^{\ast }\Psi _{j}
\end{equation}

\begin{figure}[htbp]
\centering
\includegraphics[totalheight=2.0in,trim=30 40 5 10]{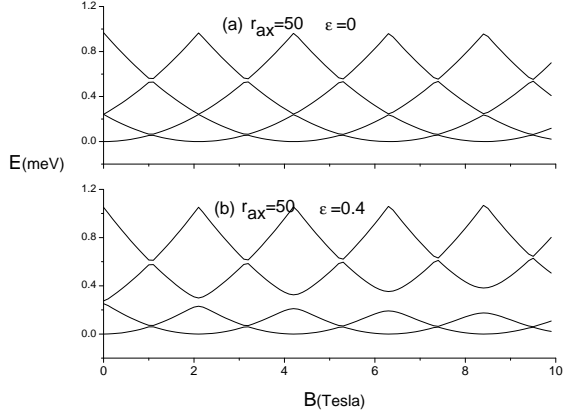}
\caption{ Low-lying spectrum (in meV)\ of an one-electron system on an
elliptic ring against $B$. \ $r_{ax}=50nm$ and $\protect\varepsilon =0\ $(a)
and 0.4 (b). The period of the flux $\protect\phi _{o}=hc/e$ is associated
with $B=2.106 \ Tesla$. }
\end{figure}

\begin{figure}[htbp]
\centering
\includegraphics[totalheight=1.8in,trim=30 20 5 10]{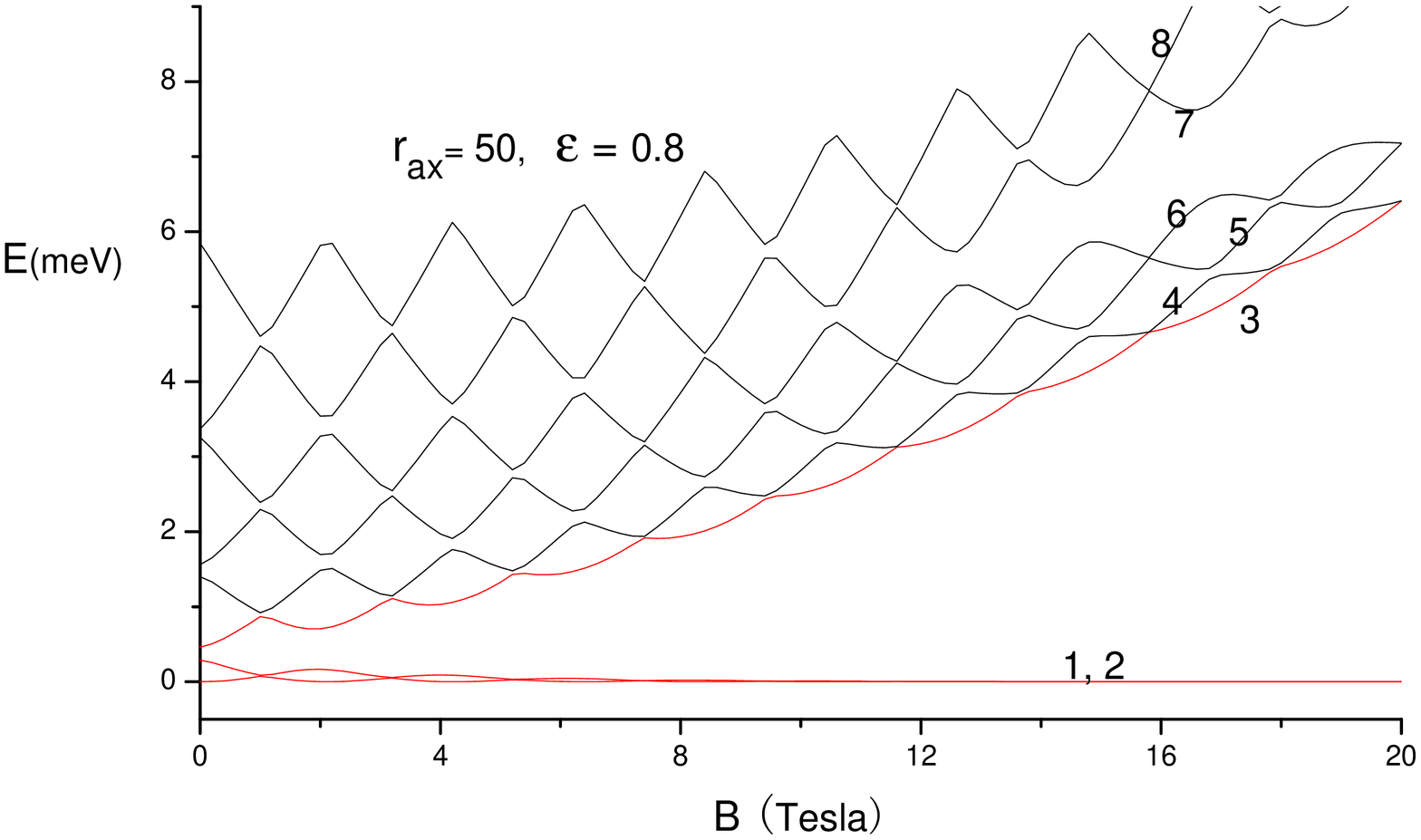}
\caption{ Similar to Fig.1 but $\protect\varepsilon =0.8$. The lowest eight
levels are included, where a great energy gap lies between the ground and
the third states. }
\end{figure}

In the follows the units meV, nm, and Tesla are used, $m^{\ast
}=0.063m_{e}$ (for InGaAs rings), and $r_{in}$ is fixed at 25. \
When $r_{ax}=50$, $ \varepsilon =0$ and 0.4, the evolution of the
low-lying spectra with $B$ are given in Fig.1. When $\varepsilon
=0.4$, the effect of eccentricity is still small, the spectrum is
changed only slightly from the case $\varepsilon =0$, but the
avoided crossing of levels can be seen.$^{10,11}$ \ In particular,
the A-B oscillation exists and the period of $\phi $ remains to be
$\phi _{o} $. \ However, when $\varepsilon $ becomes large, three
remarkable changes emerge as shown in Fig.2. (i) The A-B oscillation
of the ground state vanishes gradually. (ii) The energy of the
second state becomes closer and closer to the ground state. \ (iii)
There is an energy gap lying between the ground state and the third
state, the gap width increases nearly linearly with $B$. The
existence of the gap is a remarkable feature which has not yet been
found before from the rings with a finite width. This feature is
crucial to the optical properties as shown later. \ Fig.3
demonstrates further how the gap varies with $\varepsilon ,\
r_{ax},$ and $B$ , where $B$\ is from 0 to 30 (or $\phi $ from 0 to
14.24$\phi _{o}$). One can see that, when $\varepsilon $ is large
and $r_{ax}$ is small, the increase of $B$ would lead to a very
large gap.
\begin{figure}[tbph]
\centering
\includegraphics[totalheight=1.6in,trim=30 40 5 10]{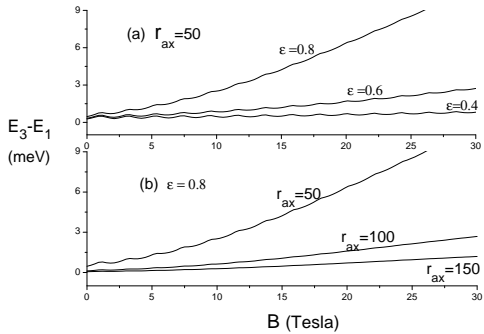}
\caption{ Evolution of the energy gap $E_{3}-E_{1}$ when $r_{ax}$
and $ \protect\varepsilon $ are given. }
\end{figure}

The A-B\ oscillation of the ground state energy is given in Fig.4. \ The
change of $\varepsilon $\ does not affect the period (2.106 Tesla). \
However, when $\varepsilon $ is large, the amplitude of the oscillation
would be rapidly suppressed. Thus, for a highly flattened elliptic ring, the
A-B oscillation appears only when $B$ is small.
\begin{figure}[htbp]
\centering
\includegraphics[totalheight=1.4in,trim=30 40 5 10]{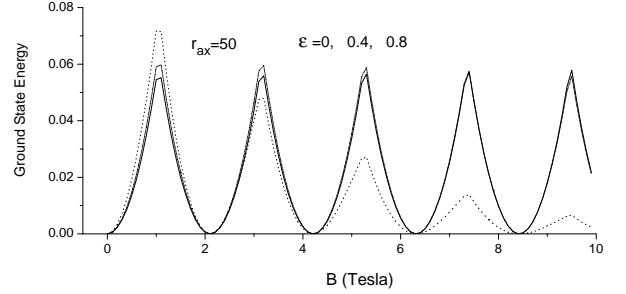}
\caption{ The A-B oscillation of the ground state energy. The solid,
dash-dot-dot, and dot lines are for $\protect\varepsilon =0,$ 0.4, and 0.8,
respectively. }
\end{figure}

The persistent current of the $j-th$ state reads$^{14}$
\begin{equation}
J_{j}=G/\hbar \lbrack \Psi _{j}^{\ast }(-i\frac{d}{d\theta }+\alpha
\frac{ \sqrt{1-\varepsilon ^{2}}}{(1-\varepsilon ^{2}\sin ^{2}\theta
)})\Psi _{j}+c.c.]
\end{equation}
The A-B\ oscillation of $J_{j}$ is plotted in Fig.5. \ When $\varepsilon $
is small ($\leq 0.4$), just as in Fig.4, the effect of $\varepsilon $ is
small as shown in 5a. When $\varepsilon $\ is large there are three
noticeable points: (i) The oscillation of the ground state current would
become weaker and weaker when $B$ increases. (ii) The current of the second
state has a similar amplitude as the ground state, but in opposite phase.
(iii) The third (and higher) state has a much stronger oscillation of
current.
\begin{figure}[tbph]
\centering
\includegraphics[totalheight=2.4in,trim=30 40 5 10]{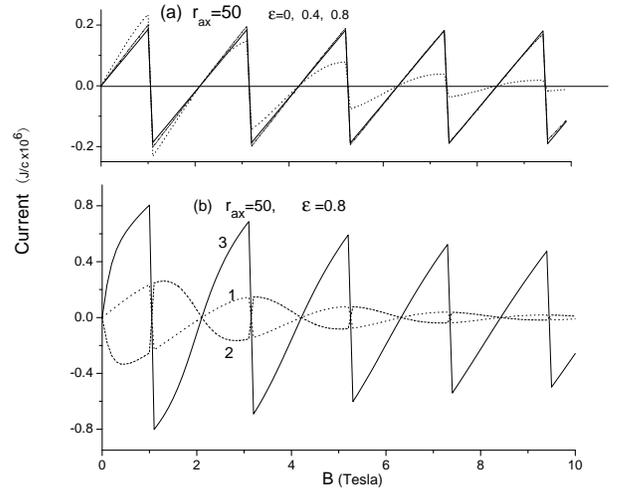}
\caption{ The A-B oscillation of the persistent current $J$. (a) is
for the ground state with $\protect\varepsilon =0$ (solid line), 0.4
(dash-dot-dot), and 0.8 (dot). \ (b) is for the first (ground),
second and third states (marked by 1,2, and 3 by the curves) with
$\protect\varepsilon $ fixed at $ 0.8$. The ordinate is $10^{6}$
times $J/c$ in $nm^{-1}$. }
\end{figure}

For elliptic rings, the angular momentum $L$ is not conserved. \
However, it is useful to define $(\overset{\_}{L})_{j}=\langle
-i\frac{\partial }{
\partial \theta }\rangle _{j}$ (refer to eq.(2)). \ This quantity would tend
to an integer if $\varepsilon \rightarrow 0$.\ It was found that (i)
When $ \varepsilon $ is small ($\leq 0.4$), $(\overset{\_}{L})_{1}$
of the ground state decreases step by step with $B$, each step by
one, just as the case of circular rings. \ However, when
$\varepsilon $ is large, $(\overset{\_}{L} )_{1}$ decreases
continuously and nearly linearly$.$\ (ii) When $\varepsilon $\ is
small, $|(\overset{\_}{L})_{i}-(\overset{\_}{L})_{1}|$\ is close
(not close) to 1 if $2\leq i\leq 3$\ (otherwise). \ Since $L$ would
be changed by $\pm 1$ under a dipole transition, the ground state
would therefore essentially jump to the second and third states.
Accordingly, the dipole photon has essentially two energies, namely,
$E_{2}-E_{1}$ and $E_{3}-E_{1}$ . However, this is not exactly true
when $\varepsilon $\ is large.

There is a relation between the dipole photon energies and the persistent
current.$^{15}$ For $\varepsilon =0$, the ground state with $L=k_{1}$ would
have the current $J_{1}=G(k_{1}+\alpha )/\pi \hbar $, while the ground state
energy $E(k_{1})=G(k_{1}+\alpha )^{2}$. \ Accordingly the second and third
states would have $L=k_{1}\pm 1$, therefore we have
\begin{equation}
|E_{3}-E_{2}|=|E(k_{1}+1)-E(k_{1}-1)|=2hJ_{1}
\end{equation}
This relation implies that the current can be accurately measured
simply by measuring the energy difference of the photons emitted in
dipole transitions. \ For elliptic rings, this relation holds
approximately when $ \varepsilon $\ is small ($\leq 0.4$), as shown
in Fig.6a. However, the deviation is quite large when $\varepsilon
$\ is large as shown in 6c.
\begin{figure}[tbph]
\centering
\includegraphics[totalheight=2.6in,trim=30 40 5 10]{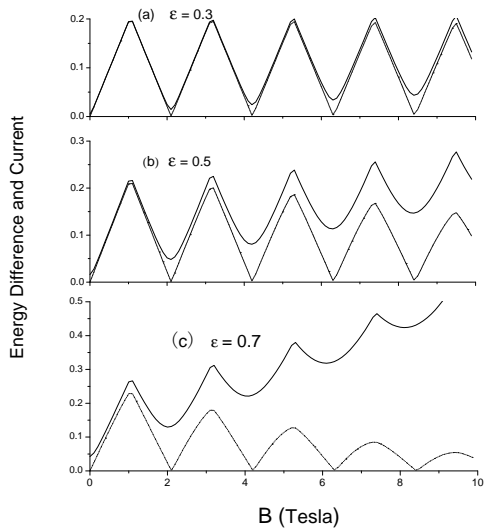}
\caption{ $E_{3}-E_{2}$ and the persistent current of the ground
state. The solid line denotes $(E_{3}-E_{2})/(2hc)10^{6}$, the
dash-dot-dot line denotes $|J|/c\cdot 10^{6}$ . They overlap nearly
if $\protect\varepsilon <0.3$. }
\end{figure}

The probability of dipole transition from $\Psi _{j}$ to $\Psi _{j^{\prime
}\ }$ reads
\begin{equation}
P_{j^{\prime },j}^{\pm }=\frac{2e^{2}}{3\hbar }(\omega /c)^{3}|\langle x\mp
iy\rangle _{j^{\prime },j}|^{2}
\end{equation}
where $\hbar \omega =E_{j^{\prime }}-E_{j}$ is the photon energy,
\begin{eqnarray}
\langle x\mp iy\rangle _{j^{\prime },j} &=&r_{ax}\int d\theta (1-\varepsilon
^{2}\cos ^{2}\theta )\Psi _{j^{\prime }}^{\ast }  \notag \\
&&[\cos \theta \mp i\sqrt{1-\varepsilon ^{2}}\sin \theta ]\Psi _{j}
\end{eqnarray}
The probability of the transition of the ground state to the
$j^{\prime }-th$ state is shown in Fig.7. When $\varepsilon $\ is
small ($\leq 0.4$) and $B$ is not very large ($\leq 10$), the
allowed final states essentially $\Psi _{2}$ and $\Psi _{3}$, and\ \
the oscillation of the probability is similar to the case of
circular rings with the same period as shown in 7a and 7b. \ In
particular, $P_{3,1}^{\pm }$ is considerably larger than
$P_{2,1}^{\pm }$ due to having a larger photon energy, thus the
third state is particularly important to the optical properties.
When $\varepsilon $\ is large (Fig.7c), the oscillation disappears
gradually with $B,$while the probability increases very rapidly due
to the factor $(\omega /c)^{3}$. Since $ E_{3}-E_{1}$ is nearly
proportional to $B$ as shown in Fig.3, the probability is nearly
proportional to $B^{3}$. This leads to a very strong emission
(absorption). Furthermore, in Fig.7c the black solid curve is much
higher than the dash-dot-dot curve, it implies that the final states
can be higher than $\Psi _{3}$, this leads to an even larger
probability.
\begin{figure}[tbph]
\centering
\includegraphics[totalheight=2.6in,trim=30 40 5 10]{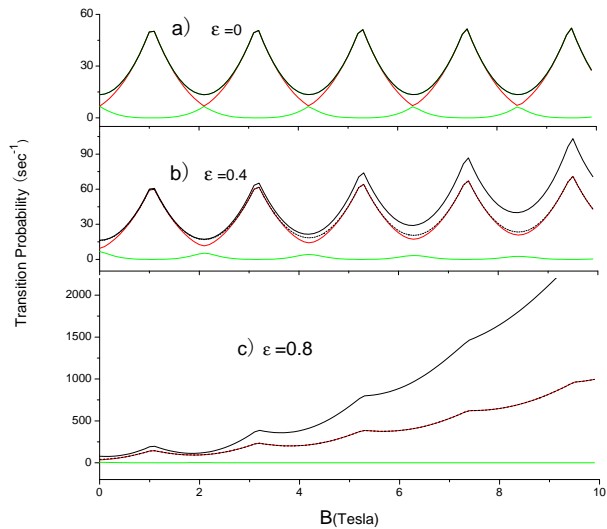}
\caption{ Evolution of the probability of dipole transition of the ground
state. The green line is for $\Psi _{1}$ to $\Psi _{2}$ transition, red line
for $\Psi _{1}$ to $\Psi _{3}$, dash-dot-dot line is for the sum of the
above two, solid line in black is for the total probability. }
\end{figure}

\begin{figure}[htbp]
\centering
\includegraphics[totalheight=3.0in,trim=30 40 5 10]{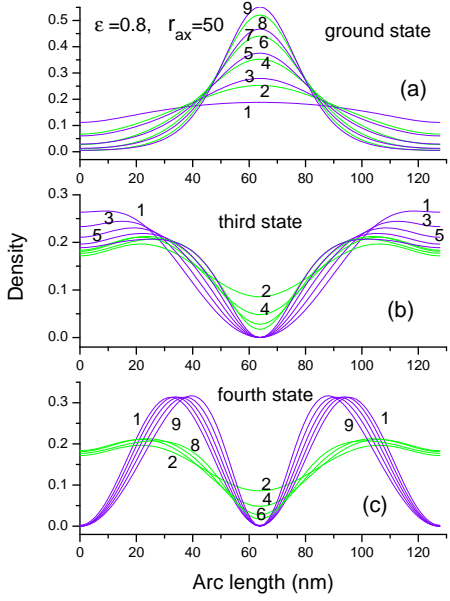}
\caption{ Particle densities $\protect\rho $ as functions of the arc
length (the according change of $\protect\theta $ is 0 to
$\protect\pi $). The fluxes are given as $\protect\phi
=(i-1)\protect\phi _{o}/2 $, where $i$ is an integer from 1 to 9
marked by the curves. The first group of curves (in violet) have
$\protect\phi /\protect\phi _{o}=$ integer, the second group (in
green) have $\protect\phi /\protect\phi _{o}=$ half-integer.
 When $\phi $
increases, the curve of $\rho $ jumps from the first group to the
second group and jumps back, and repeatedly. }
\end{figure}

For circular rings, the particle densities $\rho $ of all the
eigen-states are uniform under arbitrary $B$. \ However, for
elliptic rings, $\rho $ is no more uniform as shown in Fig.8. For
the ground state (8a), when $\phi $ =0, the non-uniformity is slight
and $\rho $ is a little smaller at the two ends of the major axis
($\theta =0,\pi $). When $\phi $ increases, the density at the two
ends of the minor axis ($\theta =\pi /2,3\pi /2$) increases as well.
When $\phi =4\phi _{o}$ the non-uniformity is very strong as shown
by the curve 9, \ where $\rho \approx 0$\ when $\theta \approx 0$ or
$\pi $. The second state has a parity opposite to the ground state,
but their densities are similar. For the third state (8b), $\rho $
is peaked not at the ends of the major and minor axes but in
between. In particular, when $ B$\ increases, $\rho $ oscillates
from one pattern (say, in violet line) to another pattern (in green
line), and repeatedly. The density oscillation would become stronger
in higher states (8c). \ The period of oscillation remains to be
$\phi _{o}$, thus it is a new type of A-B oscillation without
analogue in circular rings (where $\rho $ remains uniform).
Incidentally, the density oscillation does not need to be driven by
a strong field, instead, a small change of $\phi $ from 0 to $\phi
_{o}$\ is sufficient.

Let us evaluate $E_{j}$ roughly by using $(\overset{\_}{L})_{j}$ to replace
the operator $-i\frac{\partial }{\partial \theta }$ in eq.(2) Then,
\begin{eqnarray}
E_{j} &\approx &G\int d\theta \{[(\overset{\_}{L})_{j}+\alpha
\frac{\sqrt{
1-\varepsilon ^{2}}}{1-\varepsilon ^{2}\sin ^{2}\theta }]^{2}  \notag \\
&&+[\frac{\alpha \varepsilon ^{2}\sin 2\theta }{2(1-\varepsilon ^{2}\sin
^{2}\theta )}]^{2}\}\Psi _{j}^{\ast }\Psi _{j}
\end{eqnarray}
There are two terms at the right each is a square of a pair of
brackets (for circular rings the second term does not exist). It is
reminded that, while $ \alpha =\phi /\phi _{o}$ is given positive,
$(\overset{\_}{L})_{j}$ is negative. Thus there is a cancellation
inside the first term. Therefore, when $\varepsilon $ and $\alpha $
are large, the second term would be more important. It is recalled
that both $\Psi _{1}$ and $\Psi _{2}$ are mainly distributed around
$\theta =\pi /2$ and $3\pi /2$ (refer to Fig.8a), where the second
term is zero due to the factor $\sin 2\theta $. Accordingly the
energies of $\Psi _{1}$ and $\Psi _{2}$ are lower. \ On the
contrary, both $ \Psi _{3}$ and $\Psi _{4}$ are distributed close to
the peaks of the second term (refer to Fig.8b and 8c), this leads to
a higher energy. \ This effect would be greatly amplified by $\alpha
\varepsilon ^{2}$ , this leads to the large energy gap shown in
Fig.3.

In summary, the optical property of highly flattened elliptic narrow rings
was found to be greatly different from circular rings. \ For the latter,
both the energy of the dipole photon and the probability of transition are
low, and they are oscillating in small domains. \ On the contrary, for the
former, both the energy and the probability are not limited, the energy
(probability) is nearly proportional to $B\ (B^{3})$, they are tunable by
changing $\varepsilon ,\ r_{ax}$ and/or $B$. It implies that a strong source
of light with frequency adjustable in a wide domain can be designed by using
highly flattened, narrow, and small rings. \ Furthermore, a new type of A-B
oscillation, namely, the density oscillation, originating from symmetry
breaking, was found. This is a noticeable point because the density
oscillation might be popular for the systems with broken symmetry (e.g.,
with C$_{3}$ symmetry).

\bigskip

Acknowledgment: The support under the grants 10574163 and 90306016 by NSFC
is appreciated.

\bigskip

References

1, S.Viefers, P. Koskinen, P. Singha Deo, M. Manninen, \ Physica E \textbf{\
21 }, 1 (2004)

2, U.F. Keyser, C. F\"{u}hner, S. Borck, R.J. Haug, M. Bichler, G.
Abstreiter, and W. Wegscheider, \ Phys. Rev. Lett. \textbf{90}, 196601 (2003)

3, D. Mailly, C. Chapelier, and A. Benoit, \ Phys. Rev. Lett.\textbf{\ 70},
2020 (1993)

4, A. Fuhrer, S. L\"{u}scher, T. Ihn, T. Heinzel, K. Ensslin, W.
Wegscheider, and M. Bichler, \ Nature (London) \textbf{413}, 822 (2001)

5, A.E. Hansen, A. Kristensen, S. Pedersen, C.B. Sorensen, and P.E.
Lindelof, Physica E (Amsterdam) \textbf{12}, 770 (2002)

6, M. van den Broek, F.M. Peeters, Physica E,\textbf{11}, 345 (2001)

7, E. Lipparini, L. Serra, A. Puente, European Phys. J. B \textbf{27}, 409
(2002)

8, J. Even, S. Loualiche, P. Miska, J. of Phys.: Cond. Matt., \textbf{15},
8737 (2003)

9, C. Yannouleas, U. Landman, \ Physica Status Solidi A \textbf{203}, 1160
(2006)

10, D. Berman, O Entin-Wohlman, and M. Ya. Azbel, Phys. Rev. B \textbf{42},
9299 (1990)

11, D. Gridin, A.T.I. Adamou, and R.V. Craster, Phys. Rev. B \textbf{69},
155317 (2004)

12, A. Bruno-Alfonso, and A. Latg\'{e}, Phys. Rev. B \textbf{71}, 125312
(2005)

13, J. Goldstone and R.L. Jaffe, Phys. Rev. B \textbf{45}, 14100 (1992)

14, Eq.(4) originates from a 2-dimensional system via the following steps.
(i) the components of the current along X- and Y-axis are firstly obtained
from the conservation of mass as well known. (ii) Then, the component along
the tangent of ellipse $j_{\theta }$\ can be obtained. (iii) $j_{\theta }$\
is integrated along the normal of the ellipse under the assumption that the
wave function is restricted in a very narrow region along the normal, then
it leads to eq.(4).

15, Y.Z. He, C.G. Bao (submitted to PRB)

\bigskip

\end{document}